\documentclass[prl,twocolumn,showpacs,superscriptaddress]{revtex4}
\usepackage{graphicx}
\usepackage{amsmath}
\usepackage{bm}
\usepackage[T1]{fontenc}
\date{\today}
\begin{document}
\title{Monte Carlo simulations of ferromagnetism in p-Cd$_{1-x}$Mn$_x$Te quantum wells} 
\author{D. Kechrakos}
\affiliation{Institute of Materials Science, NCSR Demokritos,
15310 Athens, Greece}
\author{N. Papanikolaou}
\affiliation{Institute of Materials Science, NCSR Demokritos,
15310 Athens, Greece}
\affiliation{Institute of Microelectronics,  NCSR Demokritos,
15310 Athens, Greece}
\author{K. N. Trohidou}
\email{trohidou@ims.demokritos.gr}
\affiliation{Institute of Materials Science, NCSR Demokritos,
15310 Athens, Greece}
\author{T. Dietl}
\email{dietl@ifpan.edu.pl}
\affiliation{Institute of Physics, Polish Academy of
Sciences  and ERATO Semiconductor Spintronics Project, 
Japan Science and Technology Agency, al.~Lotnik\'ow 32/46, 02-668 Warszawa, Poland}

\begin{abstract}
Monte Carlo simulations, in which the Schr\"odinger equation is solved at each Monte Carlo sweep, are employed to assess the influence of magnetization fluctuations, short-range antiferromagnetic interactions, disorder, magnetic polaron formation, and spin-Peierls instability on the carrier-mediated Ising ferromagnetism in two-dimensional electronic systems. The determined critical temperature and hysteresis are affected in a nontrivial way by the antiferromagnetic interactions. The findings explain striking experimental results for modulation-doped p-Cd$_{1-x}$Mn$_x$Te quantum wells.
\end{abstract}
\maketitle
Over the recent couples of years, Mn-doped III-V and II-VI diluted magnetic semiconductors have become an important playground for developing our understanding of carrier-mediated magnetism in solids \cite{Mats02Awsc02}. To a large extend, this stems from the fact that in these systems the relevant interactions can be tuned by changing the carrier and magnetic ion densities as well as by imposing strain, confinement, electric field, or illumination.  Particularly intriguing are properties of magnetic quantum wells (QWs) \cite{Diet97,Lee00Fern01,Lee02}, as a critical dimensionality is two for a number of pertinent phenomena including the stability of ferromagnetism against magnetization fluctuations \cite{Yeom93}, Anderson localization \cite{Beli94}, spin-wave Peierls-like instability \cite{Sach02},  and formation of self-trapped magnetic polarons \cite{Kavo93Beno93}. Nevertheless, in the case of modulation-doped p-type (Cd,Mn)Te QWs, the temperature $T_{\mbox{\tiny{C}}}$ at which spontaneous spin splitting of electronic levels appears as well as its temperature dependence \cite{Haur97,Koss00,Koss02,Bouk02} follow predictions of a simple mean-field Zener-like model of ferromagnetism \cite{Diet97}. At the same time, however, wide hysteresis loops that are expected within this model \cite{Lee02} have not been observed. Instead, according to polarization-resolved photoluminescence (PL) measurements, the global spin polarization of the carrier liquid increases slowly with the external magnetic field along the easy axis, reaching saturation at a field by a factor of twenty greater than what could be accounted for by demagnetization effects \cite{Koss00,Koss02}. 

In order to determine the importance of various phenomena that control magnetism in such a reduced dimensionality disordered system, we have undertaken massive Monte Carlo (MC) simulations involving solution of the Schr\"odinger equation at each MC sweep.   Our system consists of a p-doped Cd$_{0.96}$Mn$_{0.04}$Te QW in which a magnetic order has been observed below 3~K \cite{Haur97,Koss00,Koss02,Bouk02}. The QW of width $L_W$ is sandwiched between non-magnetic (Cd,Mg,Zn)Te barriers containing acceptors that supply holes to the QW. In this modulation doping arrangement the spatial separation of the holes and the acceptors reduces significantly the importance of ionized impurity scattering.  Moreover, according to effective mass theory and optical studies of this confined system \cite{Haur97,Koss00,Koss02,Bouk02} the in-plane and perpendicular hole motions are decoupled, so that the hole wave function can be factorized as $\Psi_{\sigma}(\bm{R}) =\psi_{\sigma}(\bm{r})\varphi(z)$, where $\varphi(z) = \sqrt{2/L_W}\sin(\pi z/L_W)$. At the same time, the combined effect of confinement and biaxial epitaxial strain breaks rotational invariance, separating by 15 meV the ground state heavy hole subband $L_z = \pm 1$ from the light hole subband $L_z=0$ \cite{Koss02,Peyl93Kuhn94}. This, together with a strong spin-orbit interaction, $\lambda \bm{sL}$, where $\lambda \approx 0.7$~eV, fixes the hole spin along the $z$ axis. Accordingly, the coupling between the hole spin and Mn spin assumes an Ising-like form \cite{Haur97,Koss00,Koss02,Bouk02,Peyl93Kuhn94}, $H_{pd} = -I_{pd}(\bm{R} - \bm{R_i})s^{z}S^{z}_i$, where $I_{pd}$ is the p-d exchange function. This strong magnetic anisotropy renders phenomena associated with  quantum fluctuations of the hole spins unimportant. 

Within the standard Ruderman-Kittel-Kasuya-Yosida (RKKY) approach, the carrier-mediated exchange interaction is evaluated within the linear response theory, which is valid as long as the exchange induced spin-splitting is small compared to the Fermi energy, and the carrier wave functions are not perturbed by the Mn spins. Here, we go beyond the RKKY model, as we obtain the carrier wave functions and energies for a particular Mn spin configuration by exact diagonalization. If the p-d exchange energy $|\beta N_o| = |\langle u_k|I_{pd}|u_k\rangle|$, where $u_k$ is the Bloch amplitude normalized in the unit cell volume $1/N_o =a_o^3/4$, is greater than the width of the carrier band $W \approx 12\hbar^2/m^{*}a_o^2$, the 3D potential well associated with the single Mn atom can bind the carrier \cite{Beno92}.  Since in the case of p-Cd$_{1-x}$Mn$_x$Te, $\beta N_o = -0.88$~eV, $m^* = 0.25m_o$, $a_o=0.647$~nm, and  thus $W= 8.8$~eV, we are in the weak coupling regime, where a single Mn spin affects weakly the orbital part of the hole wave function. Hence the p-d interaction can be cast in the Fermi contact form. Thus, the Hamiltonian describing the in-plane motion of holes for a given distribution of Mn spins is $H_{KE} = p^2/2m^* - \sum_i \beta\delta(\bm{r} - \bm{r}_i)|\varphi(z_i)|^2s^zS_i^z$, where $\bm{S}_{i}$ is a classical spin vector with $S=5/2$, and $s^z$ is the hole spin density operator with $s=1/2$. We allow for antiferromagnetic (AF) interactions between Mn spins, $H_{AF} = -\sum_{i,j}k_BJ_{ij}\bm{S}_i\bm{S}_j$, with $J_{ij} = -6.3$, $-1.9$, and $-0.4$~K for the first, second, and third neighbors, respectively \cite{Shap02}. The total Hamiltonian reads $H=H_{KE}+H_{AF}$.

In our simulations we consider a finite cell $L\times L\times L_W$, cut from the (Cd,Mn)Te FCC lattice. We note that since the hole densities in the system in question are up to $10^4$ lower than that of the Mn ions we have to employ much larger cells than in previous MC simulations of III-V DMS \cite{Bose00Saka01Alva02Brey03,Schl01,Kenn02}. We take $L = 250 - 350a_o$, $L_W = 8a_o$, and  $N_S = 0.8 - 1.6\times 10^5$ Mn spins, which result in the desired Mn content $x = 0.04$. Hole concentrations in the range $p = 0.1 - 1.4\times 10^{11}$~cm$^{-2}$ are experimentally relevant and correspond to $N_h = 5 - 37$ carriers in the simulation cell \cite{note}. Periodic boundary conditions are employed in the in-plane directions and an infinite potential well is assumed normal to plane. For a given configuration of the spins, the Hamiltonian $H_{KE}$ is diagonalized in a plane-wave basis with two-dimensional (2D) wavevectors up to the truncation radius $k_c = 3.5(2\pi/L)$ \cite{Schl01}. The energy of the holes is determined by summing up the lowest $N_h$ eigenvalues of $H_{KE}$, thus neglecting carrier correlations and thermal excitations. The total energy is then obtained by adding the energy of the Mn-Mn AF interactions. Thermal fluctuations are simulated by the perturbative Monte Carlo method \cite{Kenn02}, with $2\times 10^3$ initial Monte Carlo sweeps (MCS) used for relaxation followed by $10^4$ MCS for thermal averages.

\begin{figure}
\includegraphics[width=3.1in]{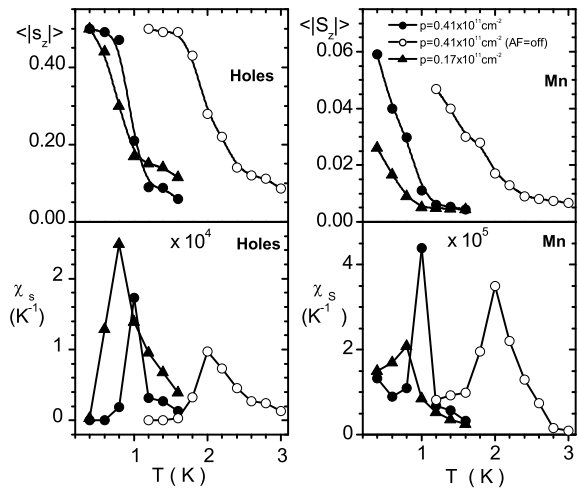}
\caption{Temperature dependence of magnetization and susceptibility of holes and Mn ions, for selected values of hole concentrations in a Cd$_{0.96}$Mn$_{0.04}$Te QW. Open circles : $J_{ij}=0$, closed circles : $J_{ij}\neq 0$.}
\label{f1}
\end{figure}

We show in Fig.~1 the temperature dependence of the spin polarization $\langle |\sigma^z|\rangle = \langle |\sum_i \sigma^z_i|/N\rangle$ and the susceptibility $\chi_{\sigma}= \left[\langle|\sum_i\sigma^z_i|^2\rangle - \langle |\sum_i\sigma^z_i|\rangle^2\right]/NT$, where $\sigma = s$ or $S$ and $N = N_h$ or $N_S$, of a Cd$_{0.96}$Mn$_{0.04}$Te QW at zero field and various hole concentrations. Non-zero values of spin polarization at high temperatures, of the order of $1/2\sqrt{N_h}$, are due to finite size effects in carrier statistics. The abrupt increase in the hole and Mn spin polarizations and the peak in the susceptibilities at the same temperature indicate a second order phase transition to a low-temperature state. While the hole spins become saturated immediately below $T_{\mbox{\tiny{C}}}$, the much denser Mn spin subsystem saturates rather slowly on lowering temperature. Notice that the range $p = 0.10 - 1.40\times 10^{11}$~cm$^{-2}$ extends from the weakly to the strongly localization regime, as evidenced by the spatial extent of the carrier wave functions, discussed below. Interestingly, spontaneous hole polarizations are observed in both these localization regimes. 

\begin{figure}
\includegraphics[width=3.3in]{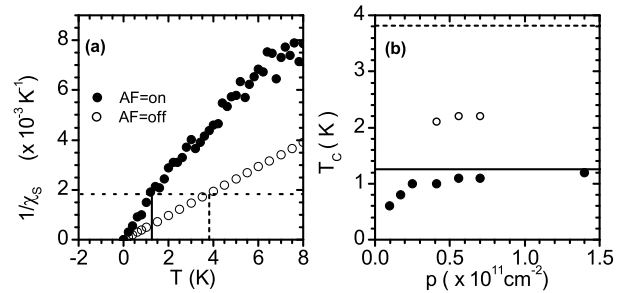}
\caption{(a) Inverse Mn susceptibility of an undoped ($H_{KE}=0$) Cd$_{0.96}$Mn$_{0.04}$Te quantum well. Open symbols :  $J_{ij} = 0$ and closed symbols : $J_{ij} \ne 0$. The intercept of the MC data and the horizontal (dotted) line at $1/\chi_S = 1.83$ (Eq.(\ref{eq2}))gives the mean-field value of the critical temperature ($T_{\mbox{\tiny{C}}}$) for the doped system. (b) Variation of $T_{\mbox{\tiny{C}}}$ with doping level as obtained by MC simulations. Horizontal lines are the mean-field results, as obtained from (a). Dashed line : $J_{ij}=0$, solid line : $J_{ij}\neq 0$.
}\label{f2}
\end{figure}
Our MC data make it possible to verify the validity of the mean-field approximation (MFA) that was previously employed to describe the experimental findings \cite{Haur97,Bouk02}. In the absence of hole scattering (disorder), the MF $T_{\mbox{\tiny{C}}}$ is determined by the condition \cite{Diet97}
\begin{equation}
A_FN_ox\beta^2\rho(\varepsilon_F)/4k_B = 1/\chi_S(T). 
\label{eq2}
\end{equation}
Here $A_F$ is the Fermi liquid parameter; $\rho(\varepsilon_F)$ is the density of the carrier states at the Fermi level, where for the 2D case $\rho(\varepsilon_F) =m^*\int \mbox{d}z|\varphi(z)|^4/\pi\hbar^2$, and $\chi_S(T)$ is the Mn spin susceptibility in the {\em absence} of carriers. Thus, within the MFA  the ferromagnetic order appears if $\chi_S^{-1}(T)< 1.83$~K, as obtained from Eq.(\ref{eq2}) for the sample in question and noninteracting holes ($A_F=1$).  Fig.~2 presents $\chi_S(T)$ as determined by MC simulations for an undoped sample ($N_h=0$), and implies that in the relevant temperature range the AF interactions reduce $\chi_S(T)$ almost threefold from the value $S^2/3T$ of the paramagnetic case, in agreement with experimental results for intrinsic Cd$_{0.96}$Mn$_{0.04}$Te \cite{Haur97,Bouk02,Shap02}. As shown in Fig.~2(a), the $\chi_{S}(T)$ data predict a  MFA value of $T_{\mbox{\tiny{C}}}$ for the doped system equal to $1.25\pm 0.1$~K and $3.81\pm 0.1$~K when the AF interactions are switched on and off, respectively. A comparison of these values to the MC results (Fig.~2(b)) demonstrates that the MFA overestimates $T_{\mbox{\tiny{C}}}$ by a factor close to two if there are no AF interactions. This could be expected given the finite range of the RKKY-type interactions involved. However, in the presence of the AF interactions the difference between the MC and MFA values of $T_{\mbox{\tiny{C}}}$ is much reduced. This implies that the system assumes the low temperature phase at significantly higher temperature than that expected for a simple ferromagnetic order.  

To elucidate further the role of AF interactions we note first that the competition between long-range ferromagnetic interactions and short-range AF interactions was suggested to result in the formation of $180^o$ domains, characterized by a wide spectrum of size distribution \cite{Koss02}. This suggestion was put forward in order to explain the absence of magnetic hysteresis and, thus, of spontaneous global magnetization in p-(Cd,Mn)Te quantum wells below $T_{\mbox{\tiny{C}}}$.  In order to investigate this conjecture we have performed MC simulations of the field dependent magnetization, shown in Fig.~3. When short-range AF interactions are switched off, the global spin polarization exhibits a square hysteresis specific to the Ising model in a magnetic field. However, in the presence of the AF interactions, a suppression of the remanent magnetization occurs and a relatively high value of the saturation field $H_{sat} \approx 150$~Oe is obtained, in agreement with the experimental findings \cite{Koss00,Koss02}. Our results indicate that the AF interactions play a role of random fields that are known to turn the hysteretic behavior of Ising spins into a linear and reversible dependence of magnetization on the external field \cite{Dahm96}.  

\begin{figure}
\includegraphics[width=3.3in]{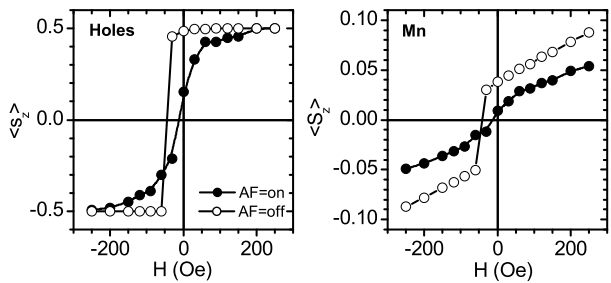}
\caption{Upper branch of the magnetization hysteresis loop of holes and Mn ions in a Cd$_{0.96}$Mn$_{0.04}$Te quantum well at $T=0.7T_{\mbox{\tiny{C}}}$. The hole density is $p=0.41\times  10^{11}$~cm$^{-2} (N_{h}=21)$.
}\label{f3}
\end{figure}
An important aspect of our simulations is that they can provide information on the actual evolution of carrier wave functions with hole density, temperature and time. We plot in Fig.~4 the in-plane charge distribution of the carrier wave functions in a Cd$_{0.96}$Mn$_{0.04}$Te quantum well with hole density $p=0.41\times 10^{11}$~cm$^{-2}$ ($N_h=21$). These snapshots correspond to the final MC sweep of simulations  carried out at high temperature ($T=5T_{\mbox{\tiny{C}}}$) and zero field. The first few ($n\simeq10$) states exhibit strong in-plane localization, with a radius $R\sim20-30$~nm, while higher states are weakly localized. Given that the one-hole states in our system are only weakly dependent on the total carrier density, we deduce  that in lightly doped samples ($p\lesssim 0.19\times 10^{11}$~cm$^{-2}$, or $N_h\lesssim 10$) all states are strongly localized while in denser samples both strongly and weakly localized states exist. 

\begin{figure}
\includegraphics[width=3.3in]{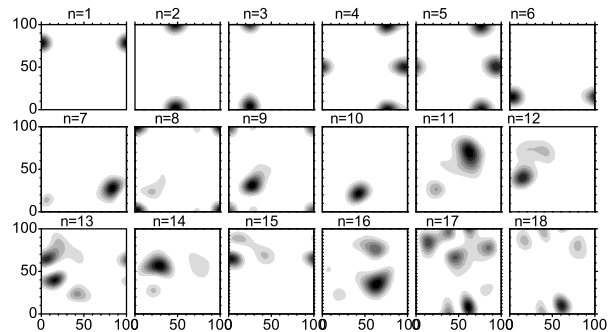}
\caption{Snapshot (at the final MCS) of the in-plane hole wave functions for the lowest eighteen eigenstates in a Cd$_{0.96}$Mn$_{0.04}$Te quantum well with $p=0.41\times  10^{11}$~cm$^{-2}$ ($N_h=21$) at high temperature ($T=5 T_{\mbox{\tiny{C}}}$). AF interactions are included. The grid scale is equal to $3.5a_o$.
}\label{f4}
\end{figure}
Furthermore, while the character of the wave functions varies appreciably with carrier density, it does not change substantially across $T_{\mbox{\tiny{C}}}$. In particular, the sequence of strongly and weakly localized states, observed above $T_c$ (Fig.~4) remains unchanged below $T_c$. This observation is consistent with the holes being scattered off the magnetic moments of Mn ions, whose ordering proceeds rather slowly on lowering temperature (Fig.~1), causing a corresponding slow variation of the scattering potential landscape. The above features compose a carrier-mediated (RKKY-type) ferromagnetic transition in a system with disorder-driven Anderson localization. As regards the time evolution of the charge distribution, at $T \ll T_{\mbox{\tiny{C}}}$ the hole gas remains fully spin polarized, while at $T \lesssim T_{\mbox{\tiny{C}}}$ our simulations indicated a slow and almost coherent reversal of the hole polarization. In this temperature regime the splitting of the PL line, being proportional to the absolute carrier polarization, will be non-zero, while its degree of circular polarization will average to zero. This behavior has indeed been observed in microluminescence studies \cite{Koss02}.

In order to assess possible effects of disorder we note that since in a true 2D system the density of states does not vary with the Fermi energy, the MFA values of $T_{\mbox{\tiny{C}}}$ are expected to be independent of the hole concentration.  However, a  lowering of $T_{\mbox{\tiny{C}}}$ when reducing $p$ has been detected experimentally in thin QWs and attributed to disorder scattering \cite{Bouk02}. This behavior is reproduced by our simulations in the low-$p$ limit (Fig.~2(b)), while the disorder-induced localization of the holes in the low-$p$ limit (Fig.~4) is consistent with the conjecture on the role of disorder on the lowering of $T_{\mbox{\tiny{C}}}$. At the same time, the lack of a substantial variation of the wave functions with temperature implies that the formation of magnetic polarons does not account for the phase transition in question. This conclusion supports the notion that the ferromagnetic instability driven by the RKKY-type interactions \cite{Diet97} precedes the self-trapping of holes and the formation of magnetic polarons \cite{Kavo93Beno93}. Finally, we note that the static spin susceptibility of the 2D carrier gas does not depend on the wave vector $q$, which suggests a tendency towards a spin-Peierls instability in such systems, occurring at $q = 2k_F$. Actually, such a spin-density wave scenario has been put forward in order to explain the above-mentioned absence of hysteresis in p-(Cd,Mn)Te quantum wells below $T_{\mbox{\tiny{C}}}$ \cite{Koss00}. If this would be the case, the spatial modulation of the carrier wave function should depend on temperature and, furthermore, its period should decrease as the square root of the carrier density. In contrast, our simulations point to an increase of the localization radius with the hole concentration.

In conclusion, we have simulated a quasi-2D system of carriers coupled to fluctuating localized spins in the regime corresponding to carrier-driven ferromagnetic instability and Anderson localization. Our results demonstrate that the RKKY-type interaction gives the dominant contribution to spin ordering, while the  formation of magnetic polarons or spin density-waves appear to be of lesser importance. Furthermore, our findings identify limits of the validity of the MFA in describing the hole-mediated ferromagnetism, particularly in the presence of competing short-range AF interactions. According to our results, the AF interactions decrease $T_{\mbox{\tiny{C}}}$ less than expected within the MFA but reduce strongly the remanence and the coercive field, which explains hitherto puzzling experimental results for p-(Cd,Mn)Te QWs. Interestingly, these effects differentiate the (II,Mn)VI from the (III,Mn)V compounds, as in the former the ionized Mn acceptors attract holes and the AF interactions, even for first neighbors, become overcompensated by the ferromagnetic coupling. Finally, it should be noted that carrier localization and spin-density wave instability in real systems are affected by the carrier-carrier repulsion, an effect neglected in the present simulations. However, we do not expect that the presence of carrier correlation will qualitative alter our conclusions.

We thank J. Cibert, D. Ferrand, J.A. Gaj,  P. Kossacki, A.H. MacDonald, and H. Ohno for valuable discussions. The work in Poland was partly supported by AMORE project(GRD1-1999-10502) of European Commission.


\begin{thebibliography}{99}
\bibitem{note} Following Ref.~\cite{Schl01}, completely filled shells in the reciprocal lattice are used, leading to $N_h=5, 9, 13, 21, 37, ...$.
\end{thebibliography}
\end{document}